\documentclass[11pt]{article}

\usepackage{amssymb,latexsym}
\usepackage{epsfig}
\usepackage{eufrak}
\usepackage{amsmath}
\usepackage{mathrsfs}
\usepackage{color}

\newtheorem{theorem}{Theorem}
\newtheorem{lemma}{Lemma}

\newcommand{\be}{\begin{equation}}
\newcommand{\ee}{\end{equation}}
\newcommand{\bee}{\begin{eqnarray*}}
\newcommand{\eee}{\end{eqnarray*}}
\newcommand{\bel}{\begin{eqnarray}}
\newcommand{\eel}{\end{eqnarray}}
\newcommand{\bec}{\begin{cases}}
\newcommand{\eec}{\end{cases}}
\newcommand{\bem}{\begin{bmatrix}}
\newcommand{\eem}{\end{bmatrix}}

\newcommand{\la}{\label}
\newcommand{\li}{\left}
\newcommand{\ri}{\right}

\newcommand{\lc}{\lceil}
\newcommand{\rc}{\rceil}
\newcommand{\lf}{\lfloor}
\newcommand{\rf}{\rfloor}

\newcommand{\lm}{\lambda}

\newcommand{\de}{\delta}

\newcommand{\al}{\alpha}

\newcommand{\f}{\frac}
\newcommand{\sq}{\sqrt}
\newcommand{\cd}{\cdots}

\newcommand{\qu}{\quad}
\newcommand{\qqu}{\qquad}
\newcommand{\fa}{\forall}

\newcommand{\mscr}{\mathscr}

\newcommand{\mrm}{\mathrm}

\newcommand{\sh}{\slash}

\newcommand{\pa}{\partial}

\newcommand{\bed}{\begin{description}}
\newcommand{\eed}{\end{description}}
\newcommand{\bei}{\begin{itemize}}
\newcommand{\eei}{\end{itemize}}
\newcommand{\ben}{\begin{enumerate}}
\newcommand{\een}{\end{enumerate}}
\newcommand{\bib}{\bibitem}
\newcommand{\beL}{\begin{lemma}}
\newcommand{\eeL}{\end{lemma}}
\newcommand{\beT}{\begin{theorem}}
\newcommand{\eeT}{\end{theorem}}

\newcommand{\bpf}{\begin{pf}}
\newcommand{\epf}{\end{pf}}
\newcommand{\bsk}{\bigskip}

\newcommand{\pfbox}{\hfill\mbox{$\Box$}}

\newenvironment{pf}{\paragraph*{Proof{\rm.}}}{\pfbox\bigskip}

\begin{document}

\title{{\bf Optimal Explicit Binomial Confidence Interval  with Guaranteed Coverage Probability}
\thanks{The author is currently with Department of Electrical Engineering,
Louisiana State University at Baton Rouge, LA 70803, USA, and Department of Electrical Engineering, Southern University and A\&M College, Baton
Rouge, LA 70813, USA; Email: chenxinjia@gmail.com}}

\author{Xinjia Chen}

\date{Submitted in April, 2008}

\maketitle

\begin{abstract}

In this paper, we develop an approach for optimizing the explicit binomial confidence interval
recently derived by Chen et al. The optimization
reduces conservativeness while guaranteeing prescribed coverage probability.

\end{abstract}

\section{Explicit Formula of Chen et al.}

Let $X$ be a Bernoulli random variable defined in probability space
$(\Omega, \mscr{F}, \Pr)$ with distribution $\Pr \{ X=1\} = 1 - \Pr
\{X=0\} =  p \in (0, 1)$.  It is a frequent problem to construct a
confidence interval for $p$ based on $n$ i.i.d. random samples $X_1,
\cd, X_n$ of $X$.

Recently, Chen et al. have proposed an  explicit confidence interval in \cite{Chen} with lower confidence limit \be L_{n, \de}  =  \frac{K}{n} +
\frac{3}{4} \; \frac{ 1 - \frac{2K}{n} - \sqrt{ 1 + \f{9} { 2 \ln \f{2}{\de} } \; K ( 1- \frac{K}{n}) } } {1 + \f{9 n} { 8 \ln \f{2}{\de} } }
\label{CI_l} \ee
 and upper confidence limit
 \be U_{n, \de}
= \frac{K}{n} + \frac{3}{4} \; \frac{ 1 - \frac{2K}{n} + \sqrt{ 1 + \f{9} { 2 \ln \f{2}{\de} } \; K ( 1- \frac{K}{n}) } } {1 + \f{9 n} { 8 \ln
\f{2}{\de} } } \label{CI_u} \ee where $K = \sum_{i = 1}^n X_i$.  Such confidence interval guarantees  that the coverage probability $\Pr \{
L_{n, \de} < p < U_{n, \de} \mid p \}$ is greater than $1 - \delta$ for any $p \in (0, 1)$.

Clearly, the explicit binomial confidence interval is conservative  and it is desirable to optimize the confidence interval by tuning the
parameter $\de$.  This is objective of the next section.

\section{Optimization of Explicit Binomial Confidence Interval}

As will be seen in Section 3, it can be shown that

\beT
 For any fixed $n$ and $p \in (0, 1)$, the coverage probability of confidence interval $[L_{n, \de}, U_{n, \de}]$ decreases as $\de$ increases.
\eeT

Hence,  it is possible to find $\de > \al$ such that
\[
\Pr \left\{ L_{n, \al} < p < U_{n, \al}  \mid p \right\} > 1 - \al,
\qqu \fa p \in (0, 1)
\]
for $\al \in (0, 1)$.  To reduce conservatism of the confidence interval, we consider the following optimization problem:

For a given $\al \in (0, 1)$, maximize $\de$  subject to the constraint that \[
 \inf_{p \in (0, 1)} \Pr \left\{ L_{n, \de} \leq p \leq U_{n, \de}  \mid p \right\} \geq 1 - \al.
 \]
A similar problem is to maximize $\de$  subject to the constraint that \[
 \inf_{p \in (0, 1)} \Pr \left\{ L_{n, \de} < p < U_{n, \de}  \mid p \right\} \geq 1 - \al.
 \]

As a result of Theorem 1, the maximum $\de$ can be obtained from $(\al, 1)$ by a bisection search. In this regard, it is essential to
efficiently evaluate $\inf_{p \in (0, 1)} \Pr \{ L_{n, \de} \leq p \leq U_{n, \de}  \mid p \}$ and $\inf_{p \in (0, 1)} \Pr \{ L_{n, \de} < p <
U_{n, \de} \mid p \}$.  This is accomplished by the following theorem derived from the theory of random intervals established in \cite{Chen2}.

\beT

Let $\de \in (0,1)$.  Define
\[
\mrm{T}^{-} (p)  =  n p + \f{ 1 - 2 p - \sq{ 1 + \f{18 n p (1-p)}{
\ln (2 \sh \de) } } } {\f{2}{3n} + \f{3}{ \ln (2 \sh \de) }} ,  \qqu
\mrm{T}^{+} (p) =
 n p + \f{ 1 - 2 p + \sq{ 1 + \f{18 n p (1-p)}{ \ln (2 \sh \de) } } } {\f{2}{3n} + \f{3}{\ln (2 \sh \de) }}
\]
for $p \in (0, 1)$ and
\[
\mscr{L}(k) = \frac{k}{n} + \frac{3}{4} \; \frac{ 1 - \frac{2k}{n} -
\sqrt{ 1 + \f{9} { 2 \ln \f{2}{\de} } \; k ( 1- \frac{k}{n}) } } {1
+ \f{9 n} { 8 \ln \f{2}{\de} } }, \qu \mscr{U}(k) = \frac{k}{n} +
\frac{3}{4} \; \frac{ 1 - \frac{2k}{n} + \sqrt{ 1 + \f{9} { 2 \ln
\f{2}{\de} } \; k ( 1- \frac{k}{n}) } } {1 + \f{9 n} { 8 \ln
\f{2}{\de} } }
\]
for $k = 0, 1, \cd, n$.  Define \[ C_l (k) =  \Pr \{ \lc \mrm{T}^{-}
(\mscr{L}(k)) \rc \leq K \leq k - 1 \mid \mscr{L}(k) \}, \qu
C_l^\prime (k) = \Pr \{ \lf \mrm{T}^{-} (\mscr{L}(k)) \rf + 1 \leq K
\leq k - 1 \mid \mscr{L}(k) \}
\]
for $k \in \{0, 1, \cd, n \}$ such that $0 < \mscr{L}(k) < 1$.
Define \[ C_u (k) = \Pr \{ k + 1 \leq K \leq \lf \mrm{T}^{+} (
\mscr{U}(k) ) \rf \mid \mscr{U}(k) \}, \qu C_u^\prime (k) = \Pr \{ k
+ 1 \leq K \leq \lc \mrm{T}^{+} ( \mscr{U}(k) ) \rc - 1 \mid
\mscr{U}(k) \}
\]
for $k \in \{0, 1, \cd, n \}$ such that $0 < \mscr{U}(k) < 1$.

 Then, the following statements hold true:

(I): $\inf_{p \in (0,1)} \Pr \{  L_{n, \de} \leq p \leq U_{n, \de}
\mid p \}$ equals to the minimum of
\[
\{ C_l (k): 0 \leq k \leq n; \; 0 < \mscr{L}(k) < 1 \} \cup \{ C_u
(k): 0 \leq k \leq n; \; 0 < \mscr{U}(k) < 1 \}.
\]

(II): $\inf_{p \in (0,1)} \Pr \{  L_{n, \de} < p < U_{n, \de} \mid p
\}$ equals to the minimum of
\[
\{ C_l^\prime (k): 0 \leq k \leq n; \; 0 < \mscr{L}(k) < 1 \} \cup
\{ C_u^\prime (k): 0 \leq k \leq n; \; 0 < \mscr{U}(k) < 1 \}.
\]

\eeT

\bsk

The proof of Theorem 2 is provided in Section 4.

\section{Proof of Theorem 1}

For simplicity of notations, define $\lm = \f{9 n} {8 \ln
\f{2}{\de}}$ and $z = \f{k}{n}$ Then, for $K = k$, the upper and
lower confidence limits are $U_{n, \de} = U(z)$ and $L_{n, \de} =
L(z)$ respectively, where
\[
U(z) = z + \frac{3}{4} \; \frac{ 1 - 2 z + \sqrt{ 1 + 4 \lm \; z  ( 1- z) } } {1 + \lm },
\qqu L(z) = z + \frac{3}{4} \; \frac{ 1 - 2 z - \sqrt{
1 + 4 \lm \; z  ( 1- z) } } {1 + \lm }.
\]
Since $(1 - 2p)^2 \leq  1 + 4 \lm \; p ( 1- p)$ for $p \in (0, 1)$
and $\lm > 0$, we have $L(z) \leq z$ and $U(z) \geq z$.  Hence, to
show Theorem 1, it suffices to show that both $U(z) - z$ and $z -
L(z)$ decrease as $\de$ increases for any $z \in [0, 1]$. We shall
first show that $U(z) - z$ decreases as $\de$ increases for any
fixed $z \in [0, 1]$.  For this purpose, we can define \[ y = \f{4
[U(z) - z]}{3}
\]
and show that $\f{ \pa y  } { \pa \lm } < 0$. To this end,  we can
use the definition of $y$ to obtain the following equation $[(1 +
\lm) y - (1 - 2z) ]^2 = 1 + 4 \lm z (1- z)$.  Differentiating both
sides of this equation with respect to $\lm$ yields
\[
2 [ (1 + \lm) y - (1 - 2 z) ] \li [ (1 + \lm) \f{ \pa y  } { \pa \lm
} + y  \ri ] = 4 z (1 - z),
\]
from which we have
\[
(1 + \lm) \f{ \pa y  } { \pa \lm } = \f{ 2 z (1 - z) } { (1 + \lm) y
- (1 - 2 z) } - y.
\]
Clearly, to show $\f{ \pa y  } { \pa \lm } < 0$,  it suffices to
show that the right-hand side of the above equation is negative for
any $z \in [0, 1]$ and $\lm
> 0$. That is, to show
\[
\f{ 2 z (1 - z)  } { \sqrt{ 1 + 4 \lm \; z  ( 1- z) }  } < y,
\]
or equivalently,
\[
(1 + \lm) 2 z (1 - z) < (1 - 2 z) \sqrt{ 1 + 4 \lm \; z  ( 1- z) }+  1 + 4 \lm \; z  ( 1- z),
\]
which can be written as $2 z (1 - z) < w(\lm)$,  where
\[
w(\lm) = (1 - 2 z) \sqrt{ 1 + 4 \lm \; z  ( 1- z) } +  1 + 2 \lm \; z  ( 1- z).
\]
Note that
\[
\f{ \pa w(\lm)  } { \pa \lm } = 2 z  ( 1- z) \li [  \f{1 - 2z } { \sqrt{ 1 + 4 \lm \; z  ( 1- z) } } + 1 \ri ]  > 0
\]
as a result of
\[
\f{1 - 2z } { \sqrt{ 1 + 4 \lm \; z  ( 1- z) } } > - \f{1} { \sqrt{ 1 + 4 \lm \; z  ( 1- z) } } > - 1.
\]
Hence, $w(\lm) > w(0) = 2(1-z)$ for any $\lm > 0$. This shows that
$2 z (1 - z) <  w(\lm)$ for any $z \in [0, 1]$ and $\lm > 0$.
Consequently,  we have established $\f{ \pa y  } { \pa \lm } < 0$,
which implies that $U(z) - z$ decreases as $\de$ increases for any
fixed $z \in [0, 1]$. Observing that $L(z) = 1 - U(1 - z)$ for any
$z \in [0, 1]$, we have
 \[ z - L(z) = z - [ 1 - U(1 - z)] = U(1 - z) - (1 - z).
\]
Therefore, it must be true that $z - L(z)$ decreases as $\de$
increases for fixed any $z \in [0, 1]$.  So, the proof of Theorem 1
is completed.

\section{Proof of Theorem 2}

For simplicity of notations, we define $\lm, \; z, \; L(z)$ and
$U(z)$ as in the proof of Theorem 1.  Note that $L(1) = 1 - \f{3}{2
(1 + \lm)} < 1$ and the derivative of $L(z)$ with respect to $z$ is
\bee L'(z) & = & 1 + \f{3}{4(1 + \lm)} \li [ - 2 - \f{1}{2} \f{4 \lm
(1 - 2 z) } { \sqrt{ 1 + 4 \lm z (1-z)
} } \ri ]\\
& = & 1 - \f{3}{2(1 + \lm)} - \f{3}{2(1 + \lm)} \f{\lm (1 - 2 z) } { \sqrt{ 1 + 4 \lm z (1-z) } }\\
& = & \f{1}{2(1 + \lm)} \li [ 2\lm - 1 - \f{ 3 \lm (1 - 2 z) } {
\sqrt{ 1 + 4 \lm z (1-z) } } \ri ] \eee which is positive if and
only if $(2 \lm - 1) \sqrt{ 1 + 4 \lm z (1-z) } > 3 \lm  (1 - 2 z)$.

To complete the proof of Theorem 2, we need some preliminary
results.

\beL \la{width} For any $n \geq 1$ and $\de \in (0,1)$,
 \be
 \la{leqpo}
\f{ (2\lm - 1)^2 ( 1 + \lm)  } { 36 \lm^2 + 4 \lm  (2\lm - 1)^2} \geq \f{1}{4}
 \ee
 if and only if $\lm \leq \f{1}{5}$.
\eeL

\bpf Note that (\ref{leqpo}) is equivalent to $(2 \lm - 1)^2 ( 1 +
\lm) \geq 9 \lm^2 +  \lm  (2 \lm - 1)^2$, which can be simplified as
$(5\lm - 1) (\lm + 1) \leq 0$.  Since $\de \in (0,1)$ and $n \geq
1$, we have $\lm > 0$.  Hence, the inequality (\ref{leqpo}) holds if
and only if $\lm \leq \f{1}{5}$.

\epf

 \beL
 $L(z)$ is monotonically increasing  with respect to $z \in [0, 1]$ such that $L(z) > 0$.
 Similarly, $U(z)$ is monotonically increasing  with respect to $z \in [0, 1]$ such that $U(z) < 1$.
\eeL

\bpf We shall first show that $L(z)$ is monotonically increasing
with respect to $z \in [0, 1]$ such that $L(z) > 0$.  It suffices to
consider four cases:

Case (i): $\lm \geq \f{1}{2}$ and $0 < z \leq \f{1}{2}$;

Case (ii): $\lm \geq \f{1}{2}$ and $1 > z > \f{1}{2}$;

Case (iii): $\lm < \f{1}{2}$ and $0 < z \leq \f{1}{2}$;

Case (iv):  $\lm < \f{1}{2}$ and $1 > z > \f{1}{2}$.

\bsk

In Case (i),  $L(z)$ increases if and only if $(2 \lm - 1)^2 [1 + 4
\lm z (1-z) ] > 9 \lm^2  (1 - 2 z)^2$, or equivalently,
\[
\li ( z - \f{1}{2} \ri )^2 < \f{ (2 \lm - 1)^2 ( 1 + \lm)  } { 36 \lm^2 + 4 \lm  (2 \lm - 1)^2  }.
\]
Define
\[
z^* = \f{1}{2} - \sq{ \f{ (2 \lm - 1)^2 ( 1 + \lm)  } { 36 \lm^2 + 4
\lm  (2 \lm - 1)^2  } }.
\]
By Lemma \ref{width}, we have $z^* > 0$. If follows that $L(z)$ is
monotonically decreasing with respect to $z \in (0, z^*)$ and
monotonically increasing with respect to $z \in \li ( z^*, \f{1}{2}
\ri )$.  This implies that $L(z)$ achieves its minimum at $z^*$ and
$L(z) < L(0) = 0$ for any $z \in (0, z^*)$.  Therefore, we have
shown that $L(z)$ is monotonically increasing with respect to $z \in
(0, 1)$ such that $L(z) \geq 0$ and that the conditions of Case (i)
hold true.

\bsk

In Case (ii), $L(z)$ increases for $z \in \li ( \f{1}{2}, 1 \ri )$.

\bsk

In Case (iii), $L(z)$ decreases for $z \in \li (0,  \f{1}{2} \ri ]$.
It can be seen that $L(z) < L(0) = 0$ for any $z \in \li (0,
\f{1}{2} \ri ]$.

\bsk

In Case (iv), $L(z)$ increases if and only if $(2 \lm - 1) \sqrt{ 1
+ 4 \lm z (1-z) } > 3 \lm  (1 - 2 z)$,  which can be written as $(1
- 2 \lm) \sqrt{ 1 + 4 \lm z (1-z) } < 3 \lm  (2 z - 1)$ or
equivalently,
\[
\li ( z - \f{1}{2} \ri )^2 > \f{ (2\lm - 1)^2 ( 1 + \lm)  } { 36 \lm^2 + 4 \lm  (2\lm - 1)^2  }
\]
Define
\[
z^\star = \f{1}{2} + \sq{ \f{ (2\lm - 1)^2 ( 1 + \lm)  } { 36 \lm^2
+ 4 \lm  (2\lm - 1)^2  } }.
\]
If $\f{1}{2} > \lm > \f{1}{5}$, by Lemma \ref{width}, we have
$z^\star < 1$.  Hence, $L(z)$ increases for $z \in (z^\star , 1)$
and
\[
L(z) < L(1) = z + \frac{3}{4} \; \frac{ 1 - 2z - \sqrt{ 1 + 4 \lm \;
z ( 1- z) } } {1 + \lm } = \frac{2 \lm - 1 } {2(1 + \lm)} < 0, \qqu
\fa z \in (z^\star , 1).
\]
Moreover, $L(z)$ decreases for $z \in \li ( \f{1}{2}, z^\star \ri )$
and
\[
L(z) < L \li ( \f{1}{2} \ri ) =  z + \frac{3}{4} \; \frac{ 1 - 2z -
\sqrt{ 1 + 4 \lm \; z ( 1- z) } } {1 + \lm } = \f{1}{2} - \f{3}{4
\sq{1 + \lm}} < 0, \qqu \fa z \in (z^\star , 1).
\]
If $0 <  \lm \leq \f{1}{5}$, by Lemma \ref{width}, we have $z^\star
\geq 1$.  Hence, $L(z)$ decreases for $z \in (\f{1}{2}, 1)$ and
\[ L(z) < L \li ( \f{1}{2} \ri ) < 0 \qqu \fa z \in \li ( \f{1}{2}, 1 \ri ).
\]

Based on the preceding investigation, we can conclude that the lower
confidence limit is non-decreasing  with respect to $z \in (0, 1)$
such that $L(z) \geq 0$. Recalling that $L(1) < 1$, we have that
$L(z) < 1$ for any $z \in (0, 1)$.

Since $U(z) = 1 - L(1-z) > 0$ for any $z \in (0, 1)$,  we have that
the upper confidence limit $U(z)$ is also non-decreasing with
respect to $z \in (0, 1)$ such that $U(z) \leq 1$.

\epf

\bsk

Now we consider the minimum coverage probability.   By the
definitions of $L_{n, \de}, \; U_{n, \de}$ and $\mscr{L}(k), \;
\mscr{U}(k)$, we have
\[
\Pr \{ L_{n, \de} \leq p < U_{n, \de} \mid \mscr{U}(k) \} =  \Pr \{
k < K \leq \mrm{T}^{+} (p) \mid \mscr{U}(k) \}, \qqu 0 < \mscr{U}(k)
< 1
\]
\[
\Pr \{ L_{n, \de} < p \leq U_{n, \de} \mid \mscr{L}(k) \} =  \Pr \{
\mrm{T}^{-} (p) \leq K < k \mid \mscr{L}(k) \}, \qqu 0 < \mscr{L}(k)
< 1
\]
\[
\Pr \{ L_{n, \de} < p < U_{n, \de} \mid \mscr{U}(k) \} =  \Pr \{ k <
K < \mrm{T}^{+} (p) \mid \mscr{U}(k) \}, \qqu 0 < \mscr{U}(k) < 1
\]
\[
\Pr \{ L_{n, \de} < p < U_{n, \de} \mid \mscr{L}(k) \} =  \Pr \{
\mrm{T}^{-} (p) < K < k \mid \mscr{L}(k) \}, \qqu 0 < \mscr{L}(k) <
1.
\]
Since both $L(z)$ and $U(z)$ are monotone, the proof of Theorem 2
can be completed by making use of the above results and applying the
theory of coverage probability of random intervals established by
Chen in \cite{Chen2}.

\end{document}